\documentstyle[preprint,prl,floats,aps,epsf]{revtex}
\begin{document}
\draft

\title{Analytical and Numerical Study of Internal Representations in 
Multilayer Neural Networks with Binary Weights}

\author{Simona Cocco \cite{sc}, R\'emi Monasson \cite{rm} 
and Riccardo Zecchina \cite{rz}}
\address{\cite{sc} INFN and Dipartimento di Fisica, P.le Aldo Moro 2,
I-00185 Roma, Italy\\
\cite{rm}  Laboratoire de Physique Th\'eorique de l'ENS,
24 rue Lhomond, 75231 Paris cedex 05, France\\ 
\cite{rz} INFN and Dipartimento di Fisica, Politecnico di Torino,
C.so Duca degli Abruzzi 24, I-10129 Torino, Italy}

\maketitle
\begin{abstract}
We study the weight space structure of the parity machine with binary weights
by deriving the distribution of volumes associated to the 
internal representations of the learning examples. The learning behaviour
and the symmetry breaking transition are analyzed and the results
are found to be in very good agreement with extended numerical simulations.
 
\end{abstract}


\narrowtext
\section{Introduction}

The understanding of the learning process of neural networks is of great
importance from both theoretical and applications points of view \cite{hertz}.
While the properties of the simplest neural network, the perceptron, are now
well explained, the picture we have for the learning phase of the far more
relevant case of multilayer neural networks remains unsatisfactory. Due to the
internal degrees of freedom present in multilayer networks (the state
variables of the hidden units), the structure of the weight space inherited
from the learning procedure is highly non trivial
\cite{hertz,nosotros,gard,bhs,bk,engel}.

Gardner's framework of statistical mechanics\cite{gard} has been proven to be
useful in understanding the learning process by providing some bounds on the
optimal performances of neural networks. In particular, it has allowed to
derive the storage capacity and the generalization abilities of neural
networks inferring a rule by example. However, the drawback of such an
approach is that it does not give any microscopic information concerning the
internal structure of the coupling space, in particular about internal
representations.

Recently, an extension of Gardner's approach has been proposed \cite{nosotros}
which leads to a deeper insight on the structure of the weight space by
looking at the components of the latter corresponding to different states of
the internal layers of the network. Such an approach has been successful in
explaining some known features of multilayer neural networks and has permitted
to find some new results concerning their learning--generalization performances
as well as to make a rigorous connection with information
theory\cite{nosotros,nosotros2}.

In this paper we focus on multilayer neural networks with binary weights
\cite{bk}. This allows us to compare the analytical study with extensive
numerical simulations and thus to provide a concrete check of the liability of
the theory. Indeed, both the structure of internal representations and the
(symmetry--breaking) learning phase transition predicted by our theory turn
out to be in remarkable agreement with the numerical findings.

The paper is organized as follows. In section 2, we present our method
from a general point of view and apply it to the parity machine with binary
weights in Section 3. Section 4 is devoted to numerical simulations.
Our results are summed up in the conclusion.

\section{Distribution of the internal representation volumes}

As discussed in ref.\cite{nosotros}, the method we adopt consists in a rather
natural generalization of the well known Gardner approach based on the study
of the fractional weight space volume not ruled out by the optimal, yet
unknown, learning process \cite{gard}.  We analyze the detailed decomposition
of such volume in elementary volumes each one associated to a possible
internal representations of the learned examples.  The dynamical variables
entering the statistical mechanics formalism are the (binary valued)
interaction couplings and the spin--like states of the hidden units.

In what follows, we focus on non--overlapping multilayer networks composed of
$K$ perceptrons with weights $J_{\ell i}$ and connected to $K$ sets of
independent inputs $\xi _{\ell i}$ ($\ell =1,\dots,K$, $i=1,\dots,N/K$).

The learning process may be thought of as a two step geometrical process
taking place in the weight space from the input to the hidden layer.  First the
$N/K$--dimensional subspace belonging to the $\ell$--th perceptron (or hidden
unit) is divided in a number of volumes ($\leq 2^P$), each of which being
labeled by a $P$--components vector
\begin{equation} 
\tau_\ell^\mu={\rm sign}(\vec{J_\ell}\vec{\xi}_\ell^\mu) \; \;, \; \; \; 
\ell=1,\dots,K \; \;, \; \; \;\mu=1,\dots,P \; \; \;.  
\label{tau}
\end{equation}
$\tau_\ell^\mu$ is the spin variable  representing the state of the $\ell$--th
hidden unit when pattern number $\mu$ is presented at the input.
Next, the solution space is defined as the direct product of the volumes
belonging to all hidden nodes and satisfying the condition imposed by the
decoder function
\begin{equation}
f\left( \{ \tau_\ell^\mu\} \right)=\sigma^\mu \; \; \;,
\end{equation}
where $\sigma^{\mu}$ is the output classifying the input pattern.  
The overall space of solution is thus composed by a set of internal volumes
$V_{\cal T}$ identified by the $K\times P$ matrix $\tau_\ell^\mu $ called {\it
internal representation} of the learning examples.  The computation of the
whole distribution of volumes $V_{\cal T}$, both their typical size and their
typical number, yields a deeper understanding on the storage problem by the
comparison of the number $\exp({\cal N}_D)$ of volumes giving the dominant
contribution to Gardner's volume with the upper bound given by total number
$\exp({\cal N}_R)$ of non--empty volumes (i.e. the total number of
implementable internal representations). Moreover, the physics of the learning
transition (the freezing phenomena and the replica symmetry breaking
transition) acquires a detailed geometrical interpretation.

Here we consider the case of Parity Machines which are characterized by a
decoder function defined as the product of the internal representation,
$\sigma^\mu=f(\{\tau_\ell\})=\prod_\ell \tau_\ell$.

As mentioned, given  a set of $P=\alpha N$ binary input--output random
relations, the learning process can be described as a geometrical selection
process aimed  to finding a suitable set of internal representations ${\cal T
} = \{ \tau _\ell ^{\mu} \}$ characterized by a non zero elementary volume
$V_{{\cal T }}$ defined by
\begin{equation}
V_{{\cal T }}  = \sum_{J_{\ell i}=\pm 1} \prod_{\mu} \theta \left(
\sigma ^{\mu} f(\{\tau_\ell ^\mu\}) \right) \prod _{\mu ,\ell} \theta \left(
\tau _\ell ^{\mu} \sum _i J_{\ell i} \xi _{\ell i} ^{\mu} \right) \; \; ,
\label{volume}
\end{equation}
where $\theta(\dots)$ is the Heaviside function. 
The overall volume of the weight space available for learning (the Gardner
volume $V_G$) can be written as
\begin{equation}
V_G=\sum_{{\cal T}}V_{{\cal T}} \; \; \;.
\end{equation}  

For the learning problem, the distribution of volumes can be derived through
the free--energy
\begin{equation}
g(r)=- \frac{1}{Nr}
\overline{\ln \left(\sum_{{\cal T}} V_{{\cal T}}
^{ {\displaystyle r}} \right) } \; \; \;,
\label{gdef}
\end{equation}
by calculating the entropy ${\cal N}\left[w(r)\right]$ of the volumes
$V_{{\cal T}}$ whose inverse sizes are equal to $w(r)=-{\frac 1N}\ln V_{{\cal
T}}$, given by the Legendre relations
\begin{equation}
w(r)={\frac{\partial (rg(r))}{\partial r}} \; \;, \; \; \;
{\cal N}\left[w(r)\right]=-{\frac{\partial g(r)}{\partial (1/r)}} \; \; \;.
\label{Legendre}
\end{equation}

When $N\to \infty $, $\frac 1N\overline{\ln (V_G)}=-g(r=1)$ is dominated by
volumes of size $w(r=1)$ whose corresponding entropy (i.e. the logarithm of
their number divided by $N$) is ${\cal N}_D={\cal N}\left[w(r=1)\right]$ and,
at the same time, the most numerous ones are those of smaller size $w(r=0)$
(since in the limit $r\to 0$ all the ${\cal T}$ are counted irrespective of
their relative volumes) whose entropy  ${\cal N}_R={\cal
N}\left[w(r=0)\right]$ is the (normalized) logarithm of the total number of
implementable internal representations. Both ${\cal N}_D$ and ${\cal N}_R$
allow to built a rigorous link between statistical mechanics and information
theory. The former (${\cal N}_D$) coincides with the quantity of information
${\cal I}=-\sum_{{\cal T}} \frac{V_{\cal T}}{V_G} \log \frac{V_{\cal T}}{V_G}$
contained in the internal representation distribution ${\cal T}$ and concerning
the weights whereas the latter (${\cal N}_R$) is the information capacity of
the system, i.e. the maximal quantity information one can extract from the
knowledge of the internal representations \cite{nosotros}.
 
\section{Analytical calculation for the Binary Parity Machine}

In the following, we shall apply the above method to derive the weight space
structure of the non--overlapping parity machine with binary couplings. The
analysis of binary models \cite{bk} is indeed more complicated than that of
their continuous counterpart due to Replica Symmetry Breaking (RSB) effects.
However, in the binary case extensive numerical simulations on finite systems
become available allowing for a very detailed check of the theory.

In the computation of $g(r)$, ${\cal N}\left[w(r)\right]$ and $w(r)$ one
assumes that, due to their extensive character, the self--averaging property
holds. We proceed in the computation of the $g(r)$ following the scheme
presented in \cite{nosotros,nosotros2} and discussed above. The basic
technical difference with the standard Gardner approach resides in the double
analytic continuation inherited from the presence of two sets of replica
indices in the weight vectors. The first coming from the integer power $r$ of
the internal volumes appearing in the partition function, the second from the
replica trick.

The replicated partition function reads 
 \begin{equation}
 \left(\sum_{ \{ \tau_\ell^\mu \} }  V_{{\cal T}}^r\right)^n=
 \sum_{ \{ \tau_\ell^{\mu\alpha} \} } \sum_{ \{ J_{\ell i}^{\alpha\nu} \}
}\prod_{\alpha,\mu}  \left[
\prod_\nu
 \left(\prod_\ell\theta\,(\tau_\ell^{\mu\alpha}\sum_iJ_{\ell
i}^{\alpha\nu}\xi_{\ell i}^\mu) \right) 
\theta(\prod_\ell\tau_\ell^{\mu\alpha})
\right] \; \; \;,
\label{Zvol}
 \end{equation}
with $\nu=1,\dots,r$ and $\alpha=1,\dots,n$ and which in turn implies the
introduction of four sets of order parameters. In the above formula, with no
loss of generality, we have posed $\sigma^\mu=1 \; \;, \; \; \; \forall \mu$.
 
At variance with Gardner's approach, the partition function (\ref{Zvol})
requires a double configuration trace, over the internal state variables and
the binary couplings.
We find
 \begin{equation}
 g(r)=- {\rm Extr}_{ Q_\ell {\hat Q}_\ell } \; \frac{1}{r}\; 
{\cal F}(Q_\ell {\hat Q}_\ell) \; \; \;,
 \end{equation}
where ${\cal F}$ reads
 \begin{eqnarray}
 & &{\cal F}({Q_\ell}\hat{Q}_\ell)=
\frac{1}{2K}\sum_\ell\,Tr(Q_\ell\hat{Q}_\ell) +  \frac{1}{K}\sum_\ell\;\ln
\left[Tr_{\{\vec{ J}_\ell\}}e^{\frac{1}{2}\vec{J}_\ell 
\hat{Q}_\ell\vec{J}_\ell}\right] + \nonumber\\
 &+&
 \alpha
\ln\left[Tr_{\{\tau_\ell^\alpha\}}\theta\,\left(\prod_\ell
\tau_\ell^{\alpha}\right) \int  \prod_\ell \frac{d\vec{x}_\ell
d\vec{\hat{x}}_\ell}{2\pi}\,\prod_{\alpha,\nu,\ell}  \theta(x_\ell^{\alpha
\nu}\tau_\ell^\alpha)\; e^{-\frac{1}{2}(\sum_\ell  \vec{\hat{x}}_\ell Q_\ell
\vec{\hat{x}}_\ell + \sum_\ell \vec{\hat{x}}_\ell \vec{\hat{x}}_\ell)  +i
\sum_\ell \vec{x}_\ell\vec{\hat{x}}_\ell}\right] \;,
\label{F}
\end{eqnarray}
with $ \vec{x}_\ell $, $\vec{\hat{x}}_\ell$, $\vec{J}_\ell$  $(n\times
r)$--dimensional vectors.
The elements of the $(n\times r)\times(n\times r)$ matrices $Q_\ell$ e $
\hat{Q}_\ell$ are the overlaps
\begin{equation}
 q_l^{\alpha,\nu_1,\beta,\nu_2}=
\frac{K}{N}\sum_i\,J_{\ell i}^{\alpha\nu_1} \,J_{\ell i}^{\beta\nu_2}
\end{equation}
between two coupling vectors belonging to the same hidden unit $\ell$
and their conjugate variables.
The simplest non trivial Ansatz (which can be physically understood within the
cavity approach \cite{rete}) on the structure of the above matrices, the
Replica Symmetric (RS) Ansatz of our approach, must distinguish elements with
$\alpha=\beta$ or $\alpha\neq\beta$, whereas ignores difference between
replica blocks and between hidden units. The matrices $Q_\ell,\hat{Q}_\ell$
become independent of $\ell$ and with elements
\begin{eqnarray}
 q_\ell^{\alpha=\beta,\nu_1,\nu_2} &=& q^* \; \;, \; \; \;
\hat{q}_\ell^{\alpha=\beta,\nu_1,\nu_2}=\hat{q}^* \nonumber \\
q_\ell^{\alpha\neq\beta,\nu_1,\nu_2} &=& q_0\; \;, \; \; \;
\hat{q}_\ell^{\alpha\neq\beta,\nu_1,\nu_2}=\hat{q_0}
\end{eqnarray}
We then find
\begin{eqnarray}
& &g(r,\hat{q}_0,q_0,\hat{q}^*,q^*)
=-\frac{1}{2}rq_0\hat{q}_0+
\frac{1}{2}(r-1)q^*\hat{q}^*+\frac{1}{2}\hat{q}^*-\nonumber \\
& &\frac{1}{r}\int\Delta x
 \ln\int\Delta y \;(2\cosh(\sqrt{\hat{q}_0}\;x
+\sqrt{\hat{q}^*-\hat{q}_0}\;y))^r- \nonumber \\
& &-\frac{\alpha}{r}\int\prod_l\Delta y_l\;
\ln[Tr_{\{\tau_l\}} \prod_{l=1}^k \int\Delta x_l\;
H(\frac{\sqrt{\hat{q}^*-\hat{q}_0}\,x_l+\tau_l \sqrt{q_0}\, y_l}
{\sqrt{1-\hat{q}^*}})^r]
\label{grs} 
\end {eqnarray}
where we have posed $Tr_{\{\tau_\ell\}} \equiv
Tr_{\{\tau_\ell\}}\theta(\prod_\ell\tau_\ell)$, $\Delta x= \exp(-x^2/2)/\sqrt
{2 \pi}$ and $H(y)=\int_y^{\infty} \Delta x$. One may notice that the above
expression evaluated for $r=1$ reduces to the RS Gardner's like  result on the
parity machine\cite{bhs} independent on the parameters $q^*$ and $\hat{q}^*$
\begin{equation}
g(r=1,\hat{q}_0,q_0,\hat{q}^*,q^*)=G_{RS}(q_0,\hat{q}_0) =
-\frac{1}{N}\overline{\ln V_G} \; \; .
\end{equation}  
where $V_G$ is the Gardner volume.
The geometrical organization of the domains is thus hidden in the Gardner
volume and shows up only when $r\neq1$ or if derivatives with respect to $r$
are considered, leading to an explicit dependence on the order parameters
$q^*, \hat{q}^*$
\begin{equation}
g(r=1+\varepsilon,\hat{q}_0,q_0,\hat{q}^*,q^*)
=G_{RS}(q_0,\hat{q}_0)+\varepsilon
\frac{\partial g}
{\partial r}\left.(\hat{q}_0,q_0,\hat{q}^*,q^*)\right|_{r=1}.
\end{equation}
In particular, the functions ${\cal N}\left[w(r=1)\right]$ and $w(r=1)$, being
derivatives of $g(r)$, will depend on $q^*$ and $\hat{q}^*$.

The RS saddle point equations read:

\noindent 1) $\frac{\partial g(r)}{\partial\hat{q}_0}=0$ :
\begin{equation}
q_0=
\int\Delta x \;
\frac{\left[\int\Delta y\; (\cosh^r(\sqrt{\hat{q}_0}\,x+
\sqrt{\hat{q}^*-\hat{q}_0}\,y))\;\tanh(\sqrt{\hat{q}_0}\,x
+\sqrt{\hat{q}^*-\hat{q}_0}\,y\right]^2}
{\left[\int\Delta y\; \cosh^r(\sqrt{\hat{q}_0}\,x+
\sqrt{\hat{q}^*-\hat{q}_0}\,y)\right]^2} \;,
\end{equation}

\noindent 2) $\frac{\partial g(r)}{\partial\hat{q}^*}=0$ : 
\begin{equation}
q^*=
\int\Delta x \;
\frac{\int\Delta y\; \cosh^r(\sqrt{\hat{q}_0}\,x+
\sqrt{\hat{q}^*-\hat{q}_0}\,y)\;\tanh^2(\sqrt{\hat{q}_0}\,x
+\sqrt{\hat{q}^*-\hat{q}_0}\,y)}
{\int\Delta y\; \cosh^r(\sqrt{\hat{q}_0}\,x+
\sqrt{\hat{q}^*-\hat{q}_0}\,y)} \;,
\end{equation}

\noindent 3) $\frac{\partial g(r)}{\partial q_0}=0$ :
\begin{equation}
{\hat q}_0 = 
\frac{\alpha K}{2 \pi(1-q^*)} \int\prod_\ell \Delta y_\ell \frac
{\left[\,Tr_{\{\tau_\ell\}}\;\tau_1 \,\prod_{\ell=2}^{K}\int \Delta x_\ell
\,H^r(A_\ell)\, \int \Delta x_1 H^{r-1}(A_1) e^{-A_1^2}\,\right ]^2 }
{\left[\,Tr_{\{\tau_\ell\}}\prod_\ell\int \Delta x_\ell
\,H^r(A_\ell)\,\right]^2} \;,
\end{equation}
in which
\begin{equation}
A_\ell=\frac{
\sqrt{q^*-q_0}\,x_\ell+\sqrt{q_0}\,\tau_\ell\,y_\ell}{\sqrt{1-q^*}} \; \; \; ,
\end{equation}

\noindent 4) $\frac{\partial g(r)}{\partial q^*}=0$ :  
\begin{equation}
{\hat q}^* =
 -\frac{\alpha K}{2\pi(1-q^*)} \int \prod_\ell \Delta y_\ell \frac
{Tr_{\{\tau_\ell\}} \,\prod_{\ell=2}^{K}\left[\int \Delta x_\ell
\,H^r(A_\ell)\right]} {\,Tr_{\{\tau_\ell\}}\prod_\ell\int \Delta x_\ell
\,H^r(A_\ell) }\, \int \Delta x_1 \;H^{r-2}(A_1)\; e^{-A_1^2}\;.
\end{equation}

The case of the parity machine is relatively simple in that a consistent
solution for the first two equations leads $ q_0=0$ and $\hat{q}_0=0$ (as it
happens in the computation of $V_G$\cite{bhs,bk}), which means that the
domains remain uncorrelated during the learning process. The latter two
equations simplify to
\begin{equation}
q^*=\frac
{\int \Delta y \cosh^r(\sqrt{\hat{q}^*}\,y)\tanh^2(\sqrt{\hat{q}^*}\,y)}
{\int \Delta y \cosh^r(\sqrt{\hat{q}^*}\,y)} \; \; \;,
\label{eq1}
\end{equation}

\begin{equation}
\hat{q}^*= \frac{\alpha K}{2\pi(1-q^*)} \;\frac
{\int\frac{dx}{\sqrt{2\pi}}\; e^{-\frac{x^2(1+q^*)}{2(1-q^*)}} \;
H^{r-2}(\sqrt{\frac{q^*}{1-q^*}}\,x)}
{\int \Delta x H^{r}(\sqrt{\frac{q^*}{1-q^*}}\,x)} \; \; \; ,
\label{eq2}
\end{equation}
with a free energy given by
  \begin{eqnarray}
  g(r,q^*,\hat{q}^*)&=&-\frac{1}{2}(1-r)q^*\hat{q}^*+\frac{\hat{q}^*}{2} 
  -\frac{1}{r}\ln\int\Delta y\;2^r\; \cosh^r(\sqrt{\hat{q}^*}\,y)+ 
 \nonumber\\
  &-&\frac{\alpha}{r}(K-1)\ln 2-\frac{\alpha}{r}K\ln\int \Delta x\; H^r(\sqrt{
\frac{q^*}
  {1-q^*}}x)\;\; \; .
\label{gsimple}
  \end{eqnarray}

For the parameters $q^*$,$\hat{q}^*$ there are two kinds of solution: a first
one $q^*=1$, $\hat{q}^*=\infty$,  which leads $w(r)=0$ and ${\cal
N}\left[w(r)\right]=(1-a)\ln2$ independently on $r$. The second kind must be
computed numerically form (\ref{eq1}) and (\ref{eq2}).

 In the replica theory, the choice of the right saddle solution, i.e. the
maximization or the minimization of the free energy, is not completely
straightforward due to the unusual $n \to 0$ analytic continuation \cite{MPV}.
Here we must deal with a double analytic continuation and the overall
criterion that must be followed is given by
\begin{eqnarray}
r< &0& \; \;, \; \; \; q_0 \to {\rm MAX}, \; q^* \to {\rm MIN} \nonumber \\
0< &r&<1 \; \;, \; \; \; q_0 \to {\rm MAX}, \; q^* \to {\rm MAX} \nonumber \\
r> &1& \; \;, \; \; \; q_0 \to {\rm MAX}, \; q^* \to {\rm MIN} \; \; \;,
\label{criterio}
\end{eqnarray}
where $MAX$ or $MIN$ indicates whether one must chose the solution which
maximizes or minimizes the free energy $g(r)$ respectively.

Like the zero entropy criterion for the binary perceptron, the behaviour of
${\cal N}[w(r)]$ and $w(r)$ (the cases $r=0$ and $r=1$ being of particular
interest) tells us when the RS Ansatz breaks down. Notice that in the binary
case also the volume size $w(r)$ assumes the role of an entropy in that it
coincides with (minus) the logarithm of the normalized number of binary weight
vectors belonging to a domain.

The Legendre transforms (\ref{Legendre}) of $g(r)$ lead to
the formulas
\begin{eqnarray}
w(r) &=& rq^*\hat{q}^* + \frac{1}{2}\hat{q}^*(1-q^*) 
  -\frac{\int\Delta y \cosh^r(\sqrt{\hat{q}^*} y)
  \ln(2\cosh(\sqrt{\hat{q}^*}\,y))}
  {\int\Delta y \cosh^r(\sqrt{\hat{q}^*} y)} - \nonumber \\
&\alpha& K\frac{\int \Delta x H^r(\sqrt{\frac{q^*}
  {1-q^*}}x)
  \ln H(\sqrt{\frac{q^*}
   {1-q^*}}x)}
  {\int \Delta x H^r(\sqrt{\frac{q^*}
  {1-q^*}}x)} \;, 
  \end{eqnarray}
and
 \begin{eqnarray}
{\cal N }\left[w(r)\right] & =&  \frac{r^2}{2}q^*\hat{q}^* + \ln \left[
\int\Delta y\;2^r\; \cosh^r(\sqrt{\hat{q}^*}\,y) \right]- \nonumber \\ &&r
\frac{\int\Delta y\; \cosh^r(\sqrt{\hat{q}^*}\,y)  
\;\ln(2\cosh(\sqrt{\hat{q}^*}\,y))}
  {\int\Delta y\;\cosh^r(\sqrt{\hat{q}^*}\,y)}
 \; + 
 \alpha (K-1)\ln2 + \nonumber \\ 
&& \alpha K \ln
\left[\int \Delta x\; H^r(\sqrt{\frac{q^*}
{1-q^*}}x)\right] 
- \alpha K r\;\frac{\int \Delta x\; H^r(\sqrt{\frac{q^*}
 {1-q^*}}x)
 \ln H(\sqrt{\frac{q^*}
  {1-q^*}}x)}
 {\int \Delta x\; H^r(\sqrt{\frac{q^*}
 {1-q^*}}x)} \; \; \;. 
\end{eqnarray}

The number ${\cal N}_D$ of domains composing $V_G$ is given by ${\cal
N}[w(r=1)]=-g(1)+w(1)$:
  \begin{eqnarray}
  {\cal N}\left[w(r=1)\right] &=& \frac{\hat{q}^*}{2}(q^*+1) -
  \frac{\int\Delta y\; \cosh(\sqrt{\hat{q}^*}\,y)
  \;\ln(2\cosh(\sqrt{\hat{q}^*}\,y))}
  {\int\Delta y\;\cosh(\sqrt{\hat{q}^*}\,y)}\; + \nonumber\\
 &+& (1-\alpha)\ln2 -
  2\alpha K \;\int \Delta x\; H(\sqrt{\frac{q^*}
  {1-q^*}}x)
  \;\ln H(\sqrt{\frac{q^*}
   {1-q^*}}x) \; \; \; . 
\label{cr1}
   \end{eqnarray}
The number ${\cal N}_R$ of the most numerous domains, i.e. the total number
of implementable internal representations, is given by the limit $r=0$.
We find 
\begin{equation}
 w(r=0)= \frac{1}{2}\hat{q}^*(1-q^*)
  -\int\Delta y
  \ln(2\cosh(\sqrt{\hat{q}^*}\,y))
  -\alpha \;K\int \Delta x\;
  \ln H(\sqrt{\frac{q^*}
   {1-q^*}}x) \; \; \;,
\label{k0}
  \end{equation}
and 
\begin{equation}
{\cal N}\left[w(r=0)\right]= \alpha (k-1)\ln2 +\alpha K \-ln \left[\frac{1}{2} +
\lim_{r \to 0} \int_{0}^{\infty}
\frac{dx}{\sqrt{2\pi}} e^{-x^2\frac{(1-q^*+rq^*)}{2(1-q^*)}}\right] \; \; \;.
\label{c0}
\end{equation}
The second term of the r.h.s. of above expression is different from zero only
if $\lim_{r \to 0} \frac{r}{1-q^*}=const.$, as it happens in the continuous
case \cite{nosotros}.
In both the continuous and binary cases, beyond a certain value $\alpha_R$ of
$\alpha$, the number of internal representations which can be realized becomes
smaller than $2^{(K-1)P}$ as the domains progressively disappear.
However, in the binary case the parameters $q^*$  does not vanish continuously
and a first order RSB transition to a theory described by two order parameters
$q^*_1,q^*_0$ is required.

At the point where the $w(r)$ vanishes the RS Ansatz must be changed.
Following the same RSB scheme as in \cite{nosotros}, the one step RSB
expression is obtained by breaking the symmetry within each elementary volume
and introducing the corresponding order parameters $(q_0^*, \hat q_0^*, q_1^*,
\hat q_1^*, m)$ in place of $(q^*, \hat q^*)$.
The free energy reads
\begin{eqnarray}
g_{RSB}(q_0&=&0,\hat{q}_0=0,q_0^*,\hat{q}_0^*,q_1^*,\hat{q}_1^*,r,m)=
\frac{1}{2}(q_1^*\hat{q}_1^* (m-1)+\hat{q}_1^* +
 q_0^*\hat{q}_0^* (r-m)) -\nonumber \\
& &\frac{1}{r}\ln \int \Delta y \left[\int \Delta z
 (2^m \cosh^m\,(\sqrt{\hat{q}_0^*}\,y + \sqrt{\hat{q}_1^*-\hat{q}_0^*}\,z)
\right]^{\frac
{r}{m}} - \nonumber\\
&& \frac{\alpha}{r}(K-1)\ln2 -
\frac{\alpha K}{r }\ln\int\Delta y \left[\int \Delta z
H^m(\frac{\sqrt{q_1^*-q_0^*}\,z +\sqrt{q_0^*}\,y)}
{\sqrt{1-q_1^*}})
\right]^{\frac{r}{m}} \; \; \;.
\end{eqnarray}
As for the binary perceptron, posing $q_1^*=1$ leads $\hat{q}_1^*=\infty$
and 
\begin{eqnarray}
g_{RSB}&(& q_0=0,\hat{q}_0=0,q_0^*,\hat{q}_0^*,q_1^*=1,\hat{q}_1^*=\infty
,m,r)\;=
\frac{1}{2}( q_0^*\hat{q}_0^*\, (r-m))\;+ q_0^*\,m)- \nonumber \\
& &\frac{1}{r}\ln \int \Delta y \;
2^{\frac{r}{m}}
 \cosh^{\frac{r}{m}}\,(\sqrt{\hat{q}_0^*}\,y\,m)
 - \frac{\alpha}{r}(K-1)\ln2 -
\frac{\alpha K}{r }\ln \int\Delta y
H^{\frac{r}{m}}  (\frac{\sqrt{q_0^*}\,y}
{\sqrt{1-q_0^*}}) \; \; \;.
\end{eqnarray}
Therefore, we may also write
\begin{equation}
g_{RSB}(q_0^*,\hat{q}_0^*,q_1^*=1,\hat{q}_1^*=\infty,m,r)\;=
\frac{1}{m} g_{RS}(\hat{q}^*= \hat{q}_0^*\,m^2, q^*=q_0^* ,r'=r/m)
\; \; .
\end{equation}
The saddle point equation with respect to $m$ reads
\begin{equation}
\frac{\partial g_{RSB}}{\partial m}=
-\frac{1}{m^2}(g_{RS}+\,r'\,\frac{\partial g_{RS}}{\partial r'}) =0\;.
\end{equation}
Such equation is nothing but the condition
\begin{equation}
w^{RS}(\hat{q}^*= \hat{q}_0^*\,m^2, q^*=q_0^* ,r'=r/m )=0 \; \; \;,
\end{equation}
that, in order to be satisfied, requires
\begin{equation}
\begin{array}{rcl}
\hat{q}_0^* & = & \frac{\hat{q}^*_c}{m^2} \; \; , \\
q_0^* & =&  q^*_c  \; \; ,\\
m & = & \frac{r}{r_c} \; \; ,
\end{array}
\end{equation}
where the parameters values  $\hat{q}^*_c$, $q^*_c$ and $r_c$ are computed
at the $w=0$ transition point.
From the relations
\begin{equation}
\frac{\partial \,r\, g_{RSB}}{\partial r}= \frac{1}{m} \frac{\partial}{\partial
 r'}\, r'\, g_{RS} \; \; ; \; \; \;
r^2\;\frac{\partial g_{RSB}}{\partial r}=r'^2\;\frac {\partial g_{RS}}{\partial
 r'} \; \; \;,
\end{equation}
it follows
\begin{equation}
w^{RSB}(r) =\frac{1}{m} w^{RS}(r_c)=0 \; \; , \; \; \; {\cal
N}^{RSB}\left[w^{RSB}(r)\right]= {\cal N}^{RS}\left[w^{RS}(r_c)\right] \; \;
\;.
\end{equation}
In Fig.1  we show the behaviour of $r g(r)$ versus $r$ for $\alpha=0.33$. The
part of the curve with positive slope cannot exists and hence beyond the $r_c$
value the function remains constant and equal to ${\cal
N}^{RS}\left[w^{RS}(r_c)\right]$.

Just like in the binary perceptron \cite{KM} or in the Random Energy Model
\cite{rem} (for which the one step RSB solution is exact), below $r_c$ and for
fixed $\alpha$, the system is completely frozen. The function $r g(r)$ behaves
like the free energy of the above mentioned systems though in such cases the
freezing takes place with  respect to the temperature and beyond the critical
temperature the free energy is equal to the constant value of the internal
energy.
The detailed phase diagram in the $\alpha \; , \; \; r$ plane is reported in
Fig.2 .

The behaviour of ${\cal N}[w(r)]$ versus $w(r)$ for $K=3$ and four different values
of $\alpha$ are shown in Fig.3 . 

One may observe four different phases:
\begin{itemize}
\begin{enumerate}

\item For $\alpha<\alpha_1=0.17$, the curve does not touch the $w=0$
abscissa and the domains have volumes between the two values
$w_1, w_2$ for which the ordinate vanishes.
For $r\;<\;r\left({\cal N}[w_2]=0\right)$ or $ r>r\left({\cal
N}[w_1]=0\right)$ the RS solution leads to a number of domains less then one
and must be rejected.
The freezing process takes place at the level of domains in that there are no
domains with $w$ values greater then $w_2$ and lower then $w_1$. The RSB
Ansatz substitutes the $q_0$ order parameter with $q_1,q_0$.

\item For $\alpha\geq 0.17$, the curve starts at $w=0$ with slope 
$r_c(\alpha)$; hence $ {\cal N}[w(r)]={\cal N}[w(r_c(\alpha))] \; ,
\;\forall\, r<r_c(\alpha)$.

\item At $\alpha=\frac{0.83}{3}=0.277$ we have $r_c(\alpha)=0$.
The value $\alpha=0.277$, where the zero temperature entropy vanishes, is
simply the critical capacity of a binary perceptron with $N/3$ input units
(the size of most numerous domains corresponds to the solution volume of a
subperceptron).
Beyond this $\alpha$ value, the curve will be enclosed in the region of
positive slope ($r \geq 0$) and the number of internal representations ${\cal
N}_R$ it is no longer $2\alpha\ln2$ (i.e. the maximal one) but is given by the
value of ${\cal N}[w(r)]$ at the starting point of the curve:
\begin{equation}
{\cal N}_R ={\cal N}[w(r_c(\alpha))] \; \; \;.
\end{equation}

\item At $\alpha =0.41$  the starting slope is $r_c(0.41)=1$ and
${\cal N}[w(r)=0]=(1-\alpha)\ln2$ (consistent with the condition $ g(1)=(1-\alpha)\ln 2$).

\item For $\alpha>0.41$, the point $ {\cal N}[w(r)]=(1-\alpha)\ln2 $ is off
the curve and $r_s(\alpha)$ is the point at which the two solutions of the
saddle point equations lead to the same free energy value, i.e. such that
\begin{equation}
-\frac{{\cal N}[w(r_s(\alpha)]}{r_s(\alpha)}+w(r_s(\alpha)) =-\frac{1}{r_
s(\alpha)}(1-\alpha)\ln2 \; \; \;.
\end{equation}
The starting point of the curve $(r_s(\alpha))$ grows with $\alpha$. For $ r <
r_s(\alpha)$, the correct saddle point solution is the one giving ${\cal
N}[w(r)]=(1-\alpha)\ln2$ independently  on $r$, i.e. the isolated point marked
in Fig.3 . The switch between the two solutions can be understood by noticing
that it correspond to the only possible way of obtaining $g(1)=(1-\alpha) \ln2
$ for $\alpha<0.41$. Moreover, its physical meaning is that for $r<r_s(\alpha)$
it is not necessary to distinguish among different domains in that $V_G$ is
dominated by the domains of zero entropy independently on the freezing process.
 
\item For $\alpha=0.56$ only one point remains.

\item At $\alpha=1$ also the point disappears.

\end{enumerate}
\end{itemize}

In the following section we will compare the behaviour of ${\cal N}_R$ and
${\cal N}_D$ computed for $K=3$ with the results of numerical simulations on
finite systems.

Very schematically we have 
\begin{equation}
{\cal N}_R=\left\{   \begin{array}{lc}
                   2\,\alpha(\ln2) &\alpha\leq 0.277  \\
                   {\cal N}\left[w(r_c(\alpha))\right]  & 0.277<\,\alpha<\,0.41 \\ 
                   (1-\alpha)\ln2  & \alpha>0.41
             \end{array}
    \right.
\end{equation}
and
\begin{equation}
{\cal N}_D=\left\{
\begin{array}{lc}
{\cal N}\left[w(r)=1\right] & \alpha \leq 0.41 \\
(1-\alpha)\ln2 & \alpha \geq 0.41 \end {array}\right.
\; \; \;.
\end{equation}

The overall scenario arising from the analytical computation may be summarized
briefly as follows. We find a freezing transition at  $\alpha_2=0.41$ within
the domains. For values of $\alpha>\alpha_2$ the domains, though still
distributed over the whole space of solution ($q_0=0$), are composed by
configurations with overlap $q^*=1$. The point ${\cal N}_D=0$ is the symmetry
breaking point also corresponding to the critical capacity of the model
$\alpha_c=1$ \cite{bk}.

\section{Numerical Simulations}

We have checked the above scenario by performing two distinct sets of extended
numerical simulations on the weight space structure of a parity machine with
binary weights and three hidden units.

In the first simulation we have measured both the dimension $w(r)$ and the
number ${\cal N}[w(r)]$ of domains depending on the loading parameter
$\alpha$. In particular we have considered the cases $r=1$ and $r=0$ giving
respectively the measure of  the number ${\cal N}_D$ of domains contributing
to the total Gardner volume $V_G$ and the overall number ${\cal N}_R$ of
implementable internal representation. In the second set of simulations we
have reconstructed the plot  of $rg(r)$ and ${\cal N}[w(r)]$ as function of
$r$ and for fixed $\alpha$.

The numerical method adopted is the exact enumeration of the configurations
$\{J_{\ell i} \}$ on finite systems. Very schematically the procedure is the
following.
\begin{itemize}
\begin{enumerate}
\item choose $P$ random patterns;

\item divide, for every subperceptron, the set of $2^n \;(n=N/3)$
configurations in subsets labeled by the vectors $\vec{\tau_\ell}$
($\tau_\ell^{\mu}=sign(\vec{J_\ell}\cdot\xi_\ell^{\mu})$) $\ell=1,2,3$

\item try all the subsets combinations between the three subperceptron
and identify the domains of solutions as those which
satisfy  $\prod_\ell \tau_\ell^{\mu}=1\, , \forall \mu$.

\end{enumerate}
\end{itemize}

The above scheme yields a parallel enumeration and classification of the $2^n$
weights configurations in the three subperceptron.
To avoid ambiguities in the signs of the hidden fields the number of inputs
connected to each hidden unit must be odd. The sizes of the systems taken
under consideration are $N=15, 21, 27$ for the first type of simulation and
$N=15, 21, 27, 33$ for the second.
 
More in detail, the three steps of the numerical procedure are the following. 
\begin{itemize}
\begin{enumerate}

\item We use Gaussian patterns in order to reduce finite sizes
effects (as has been done for the binary perceptron \cite{gard,KM,GS,KO}). 
From the replica method one expects
that the results are equivalent to those of binary weights in that
they depend only on the first
two moments of the quenched variables.

\item The classification of the $2^n$ weights configurations is 
as follows:
we start with $\vec{J}=(-1,-1,\ldots,-1)$. Next we compute for every $\ell$
and $\mu$ the field $a_\ell^{\mu}=-\sum\xi_{\ell i}^{\mu}$ together with its
sign $(\tau_\ell^{\mu})$ so that the vector $\vec{\tau_\ell}$ labels the first
subset. The subsequent $J$ configurations are generated by means of the Gray
code which flips just one of the $J_i$ components at each time step and allows
to update the field values with a single operation $a_{\ell}^{\mu}=
a_{\ell}^{\mu}+2\xi_{\ell i}^{\mu}$ (this reduces the number of operations by
a factor n). Then, depending on whether the  vector $\vec{\tau_\ell}$ is
different from the previous one or not, we use $\vec{\tau_\ell}$ as new label
of the second subset or increment the number of vectors contained in the first
one. We thus proceed in this way to scan the $\vec{J}$ configurations. If $P$ 
varies from 1 to $3n$, every $\vec{J}$ configuration is classified $n$ times
on each subperceptron. At the end we obtain $3$ ($P$ fixed) or $3n$ (P varying
from 1 to 3n) tables whose columns (in number $\leq 2^{n}$) are the
$\vec{\tau_\ell}$ vectors labeling the subsets and to which are associated the
numbers of $J_\ell$ belonging to each subset.

\item Finally, in the case of a given $P$,  we take a column in each of
the three tables and verify 
whether the product between
the two chosen  columns from the first two tables is equal to the column
of the third one. If so, the internal representation given by the
three columns matrix is implementable and the volume of the corresponding domain
is the product of the numbers of $J_\ell$ belonging to the subset. 

\end{enumerate}
\end{itemize}

Once the domains volumes $(V_{\cal T})$ have been measured, we compute:
\begin{equation}
-rg(r)=\ln \sum_{\cal T} V_{\cal T}^{r} \; \; \;,
\end{equation}
\begin{equation}
-w(r)=\frac{\sum_{\cal T} V_{\cal T}^{r} \ln V_{\cal T}}{\sum_{\cal T} V_{\cal
T}^{r} }
\; \; \; ,
\end{equation}
(which is the domain size computed on the saddle point of the partition
function) and
\begin{equation}
{\cal N}[w(r)]=-r g(r)+r w(r)
\; \; \;.
\end{equation}

For the first set of simulations, the above functions are computed just for
$r=0, 1$ and  the averages are taken over 10000  (N=15), 1000 (N=21) or 50
(N=27) samples.  In the case of the second set of simulations, in order to
allow for a comparison between all the finite sizes considered, $\alpha$ is
settled at $\alpha=0.33$. $r$ runs from -1.5 to 3 and the average is done over
10000 (N=15,N=21), 5000 (N=27) or 200 (N=33) samples. The statistical errors
bars are within $0.1 \%$.

As shown in Fig.4 , both theoretical and experimental results give
$g(r=1)=-(1-\alpha)\ln2$ which coincides with the annealed approximation
(so  that the total volume  is reduced simply to a half for every
added pattern and $\alpha_c=1$ \cite{bhs}). 
At the  value $\alpha=0.277$ (Fig. 5), the total number of internal state
vectors belonging to the most numerous volumes (i.e. volumes characterized by
$r=0$) becomes non--extensive ($w(r=0)=0$). Beyond such a value and in perfect
agreement with simulations, the correct solution is given by  one step of RSB
which, in fact, predicts $w(r=0)=0 \; \; , \; \forall \alpha >0.277$.

As shown in Fig.6, beyond $\alpha=0.27$ the domains begin to disappear and the
number of internal representations ceases to be constant (equal to
$2\alpha\ln2$) and starts to decrease with $\alpha$. For $r=1$ the freezing
transition takes place at $\alpha=0.41$, see Fig.7 and Fig.8 .

As shown in Fig.1, for $\alpha=0.33$ the theoretical value for the freezing
transition is $r_c=0.4$; for $r<r_c$ the slope of the curve  $rg(r)$ is zero
(it cannot become positive) and $rg(r)=-{\cal N}[w(r_c)]=-0.43$. Finally, the
plot of  ${\cal N}[w(r)]$ versus $w(r)$, for $\alpha=0.33$, is given in Fig.9.

\section{Conclusion}

In this paper we have applied the internal representation volumes approach  to
the case of binary multilayer networks, in particular to the non--overlapping
parity machine. The chief result of our study consists in a detailed
comparison between the analytical prediction and the numerical simulations,
allowing for a definitive confirmation of the method. The detailed geometrical
structure of the weights space predicted by the theory, both ${\cal
N}_D$--${\cal N}_R$ as well as the RSB transitions within the volumes, turn
out to be in remarkable agreement with the numerical simulations performed  on
finite systems.

As a general remark, let us emphasize that multilayer neural networks with
binary weights  behave differently from their continuous counterpart. While
the breaking of symmetry in the former occurs inside the representations
volumes, we have already shown that in the case of real valued couplings the
transition takes place between different volumes \cite{nosotros}. Therefore,
the richness of the distribution of internal representations found in the
continuous case, i.e. the presence of a ``finite'' number of macroscopic
regions in the weight space containing a very large number of different
internal representations, is partially lost when one deals with discrete
weights.

The method can be easily extended \cite{nosotros,nosotros2} to address the
rule inference capability problem. Thus, another very interesting and
important issue related to the present approach would be the study of the
distribution of metastable states arising from a gradient learning process.
Work is in progress along these lines.

\vfill
\eject

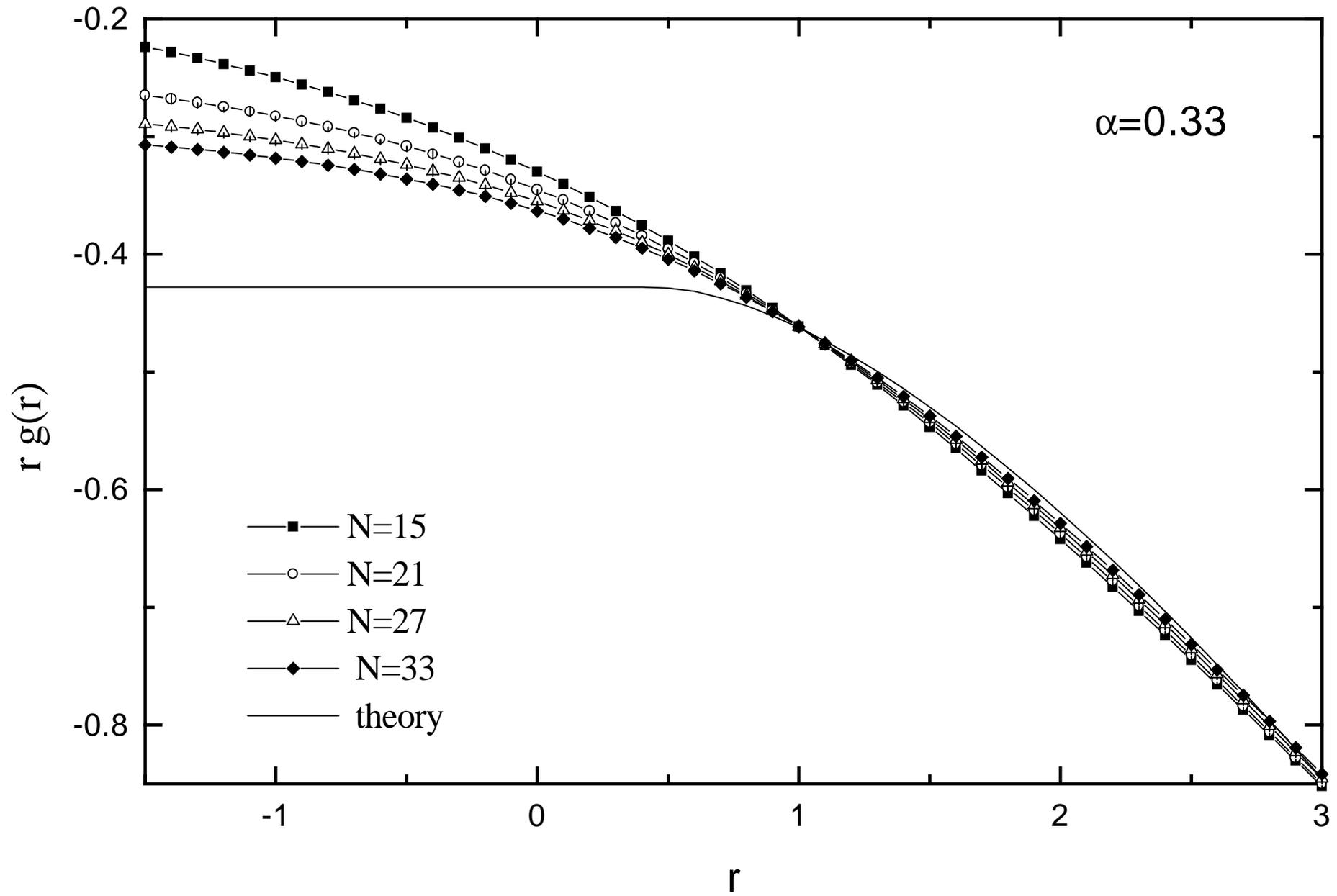
\begin{figure}
\caption{$rg(r)$ versus $r$ for $\alpha=0.33$ and $K=3$. The theoretical curve 
corresponds to the continuous lines whereas the marked curves
are the numerical results obtained for $N=15, 21, 27, 33$.}
\label{fig1}
\end{figure}

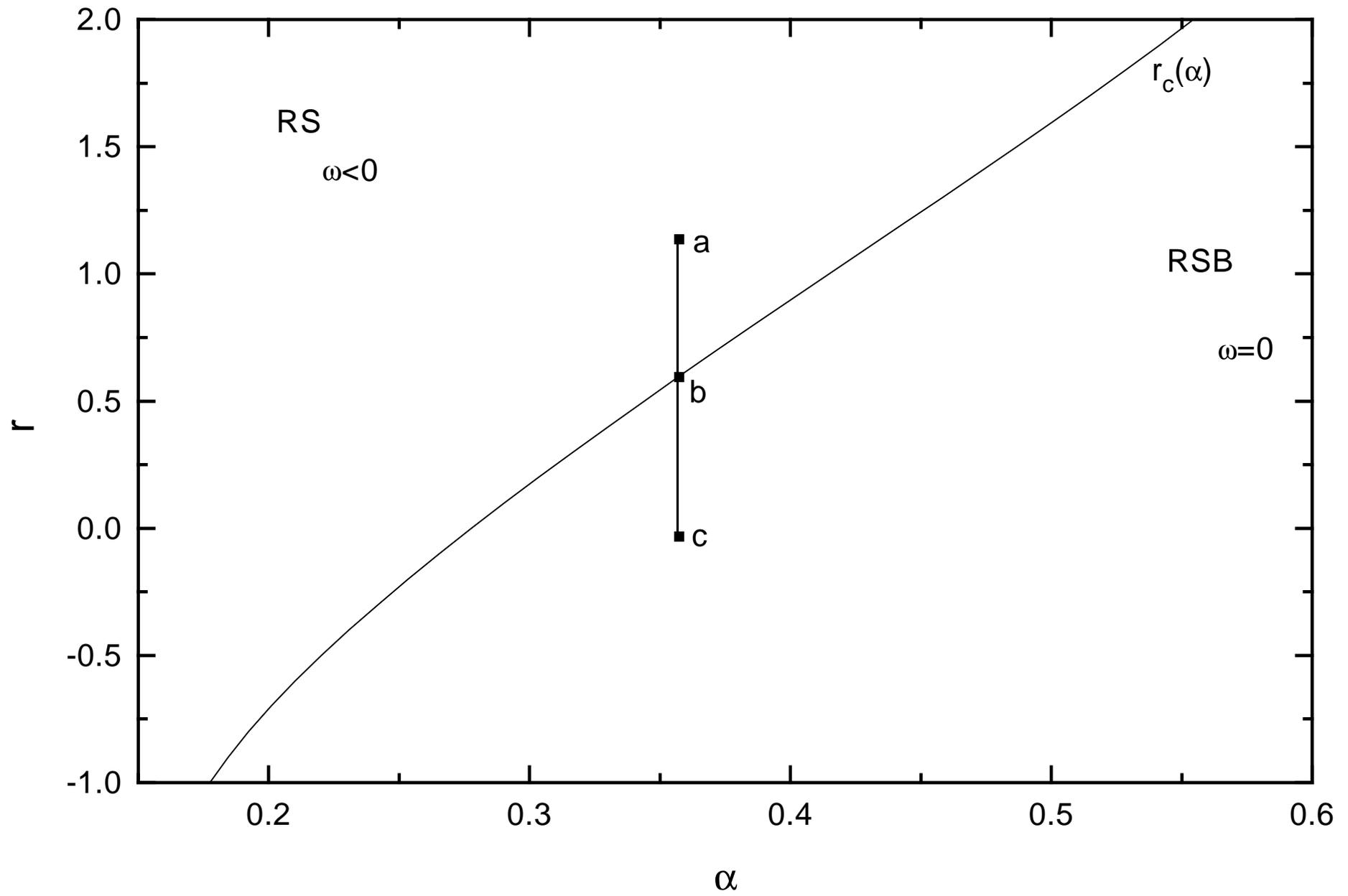
\begin{figure}
\caption{Freezing transition for the binary parity machine for $K=3$.
The $r_c(\alpha)$ line separates the RS and the RSB phases.
The three marked points describe the transition at $\alpha={\rm const.}$
and correspond to the following values of the parameters and the entropy: (a)
$q^*(r), \hat q^*(r), {\cal N}[w(r)]$, (b) $q^*(r_c), \hat q^*(r_c), {\cal
N}[w(r_c)]$ and (c) $\hat q_0^*=\hat q^*(r_c)/m^2$, $q_0^*=q^*(r_c)$,
$q_1^*=1$, $\hat q_1^*=\infty$, $m=r/r_c$, ${\cal N}[w(r)]={\cal N}[w(r_c)]$.}
\label{fig2}
\end{figure}

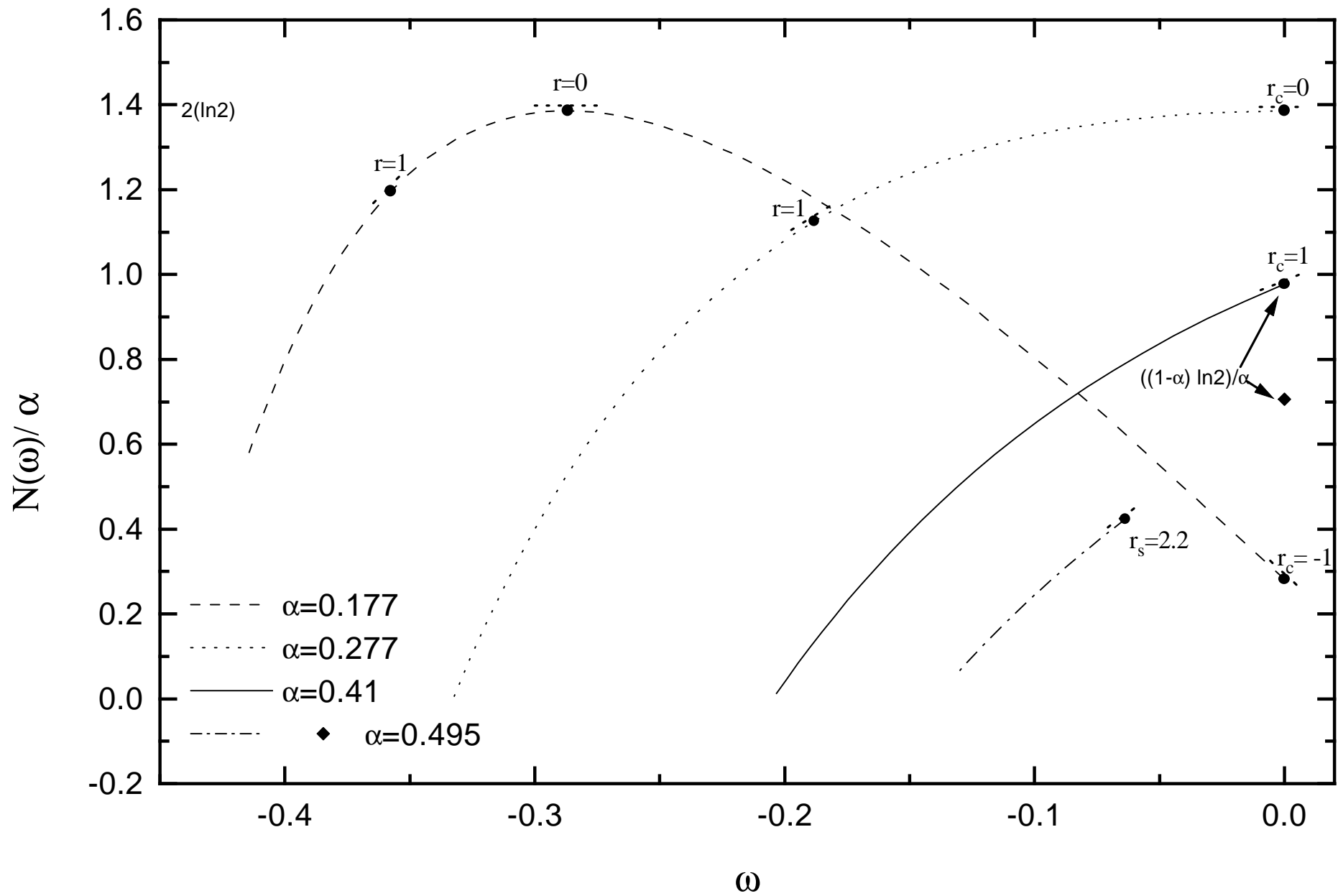
\begin{figure}
\caption{${\cal N}[w(r)]/\alpha $ versus $w(r)$ for $\alpha=0.177, 0.277,
0.41, 0.495$. The dotted points signal the starting points ($r_c$)
corresponding to $w(r_c)=0$ and the points with slope $r=0$ and $r=1$. Notice
that the diamond bolded point belongs to the dashed--dotted curve.}
\label{fig3}
\end{figure}

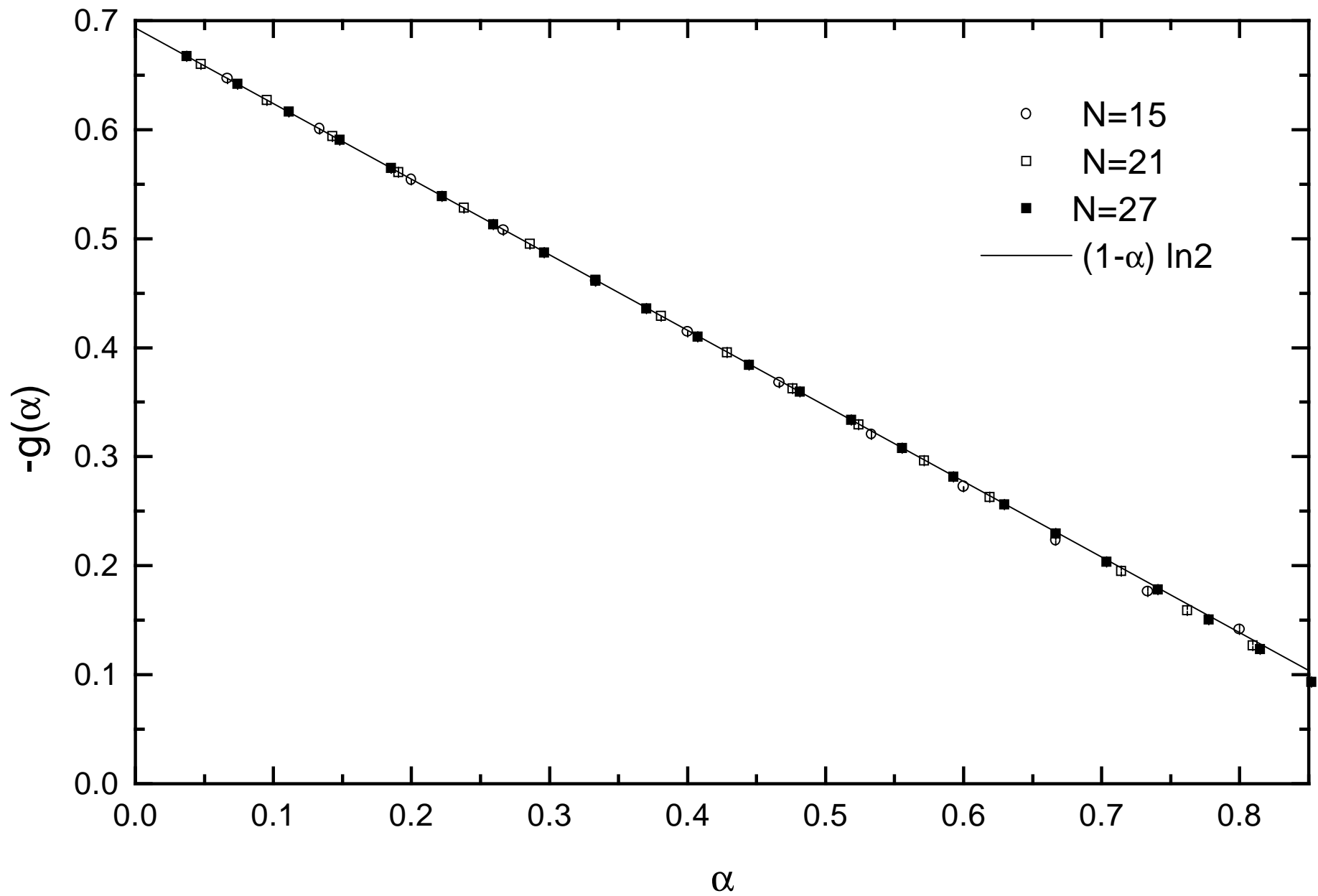
\begin{figure}
\caption{$g(r=1)$ versus $\alpha$. The theoretical curve (continuous line)
is compared with the numerical outcomes (marked points).}
\label{fig4}
\end{figure}

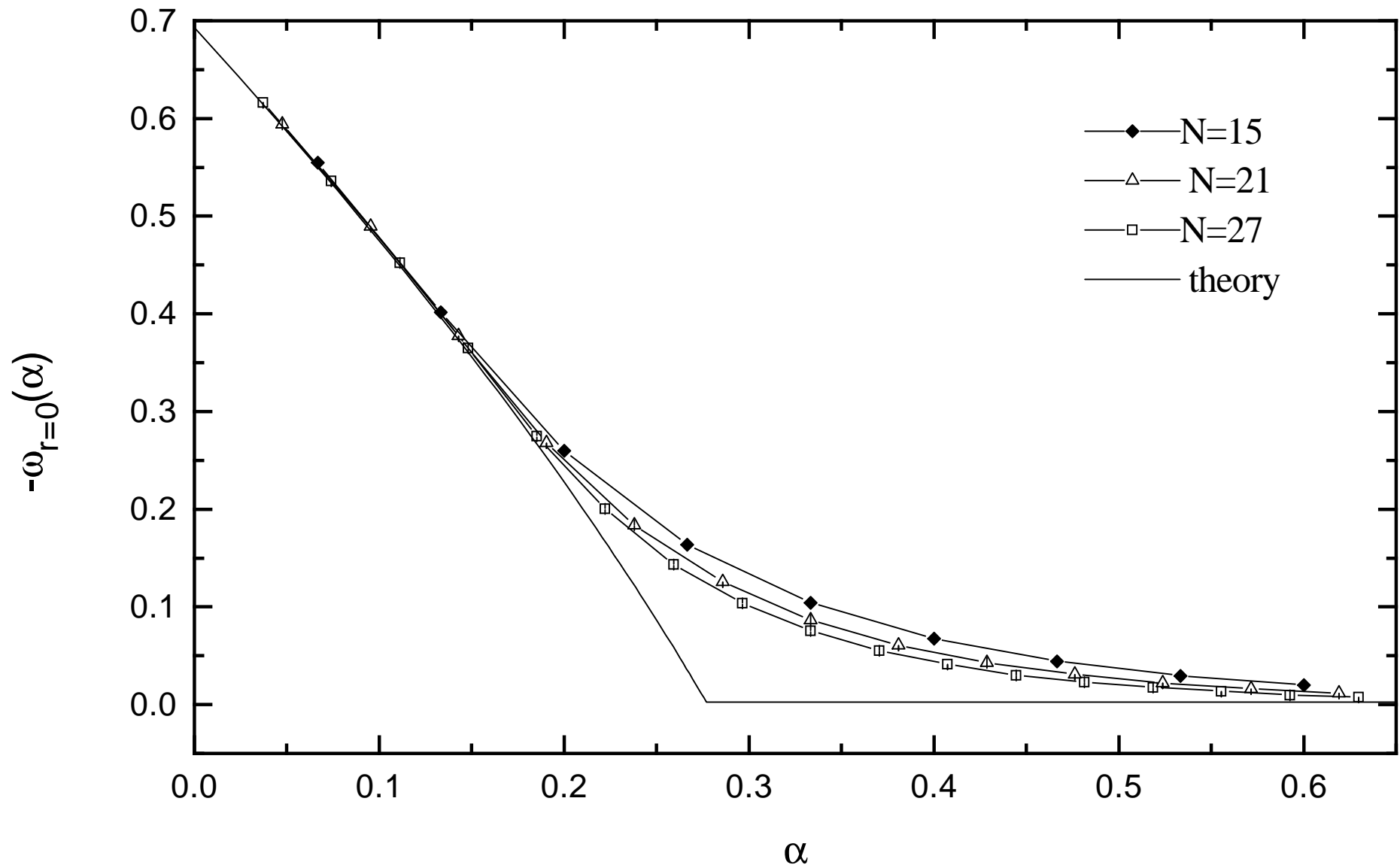
\begin{figure}
\caption{$-w(r=0)$ versus $\alpha$ (theoretical continuous line and numerical
points). The $r=0$ freezing transition appears at $\alpha=0.277$.}
\label{fig5}
\end{figure}

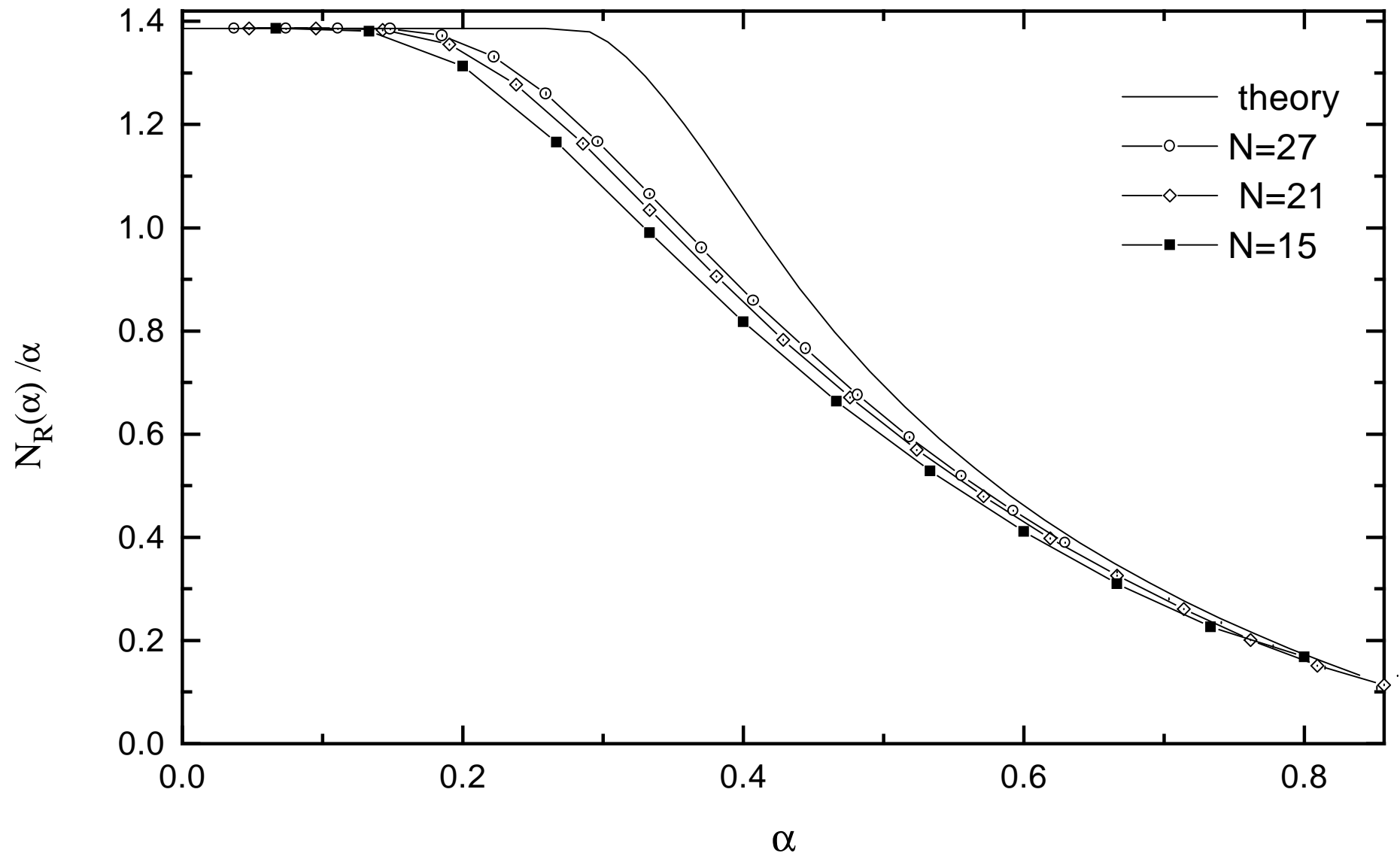
\begin{figure}
\caption{${\cal N}_R/\alpha$ versus $\alpha$ (theoretical continuous line and
numerical points).}
\label{fig6}
\end{figure}

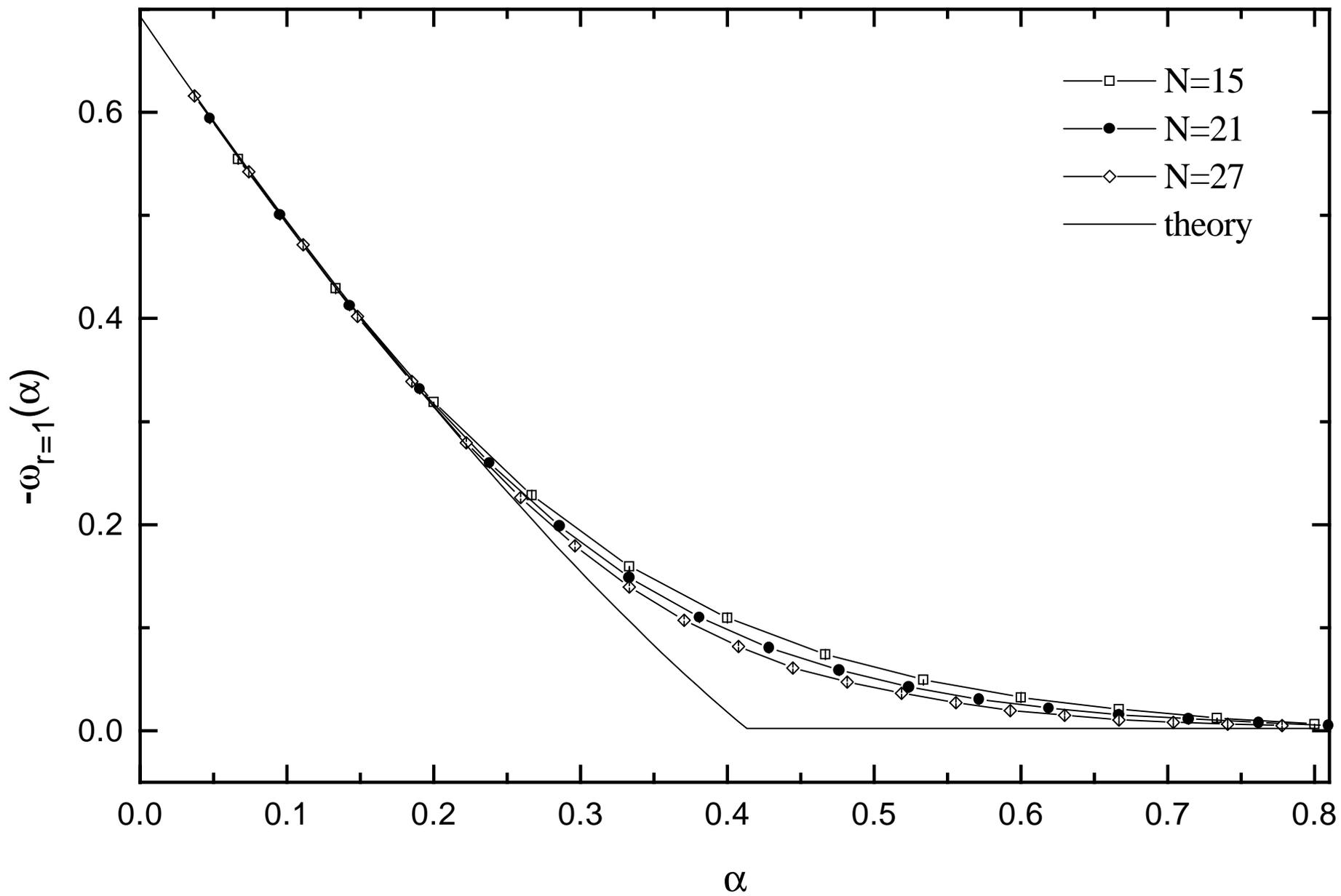
\begin{figure}
\caption{$-w(r=1)$ versus $\alpha$ (theoretical continuous line and numerical
points). The $r=1$ freezing transition appears at $\alpha=0.41$.}
\label{fig7}
\end{figure}

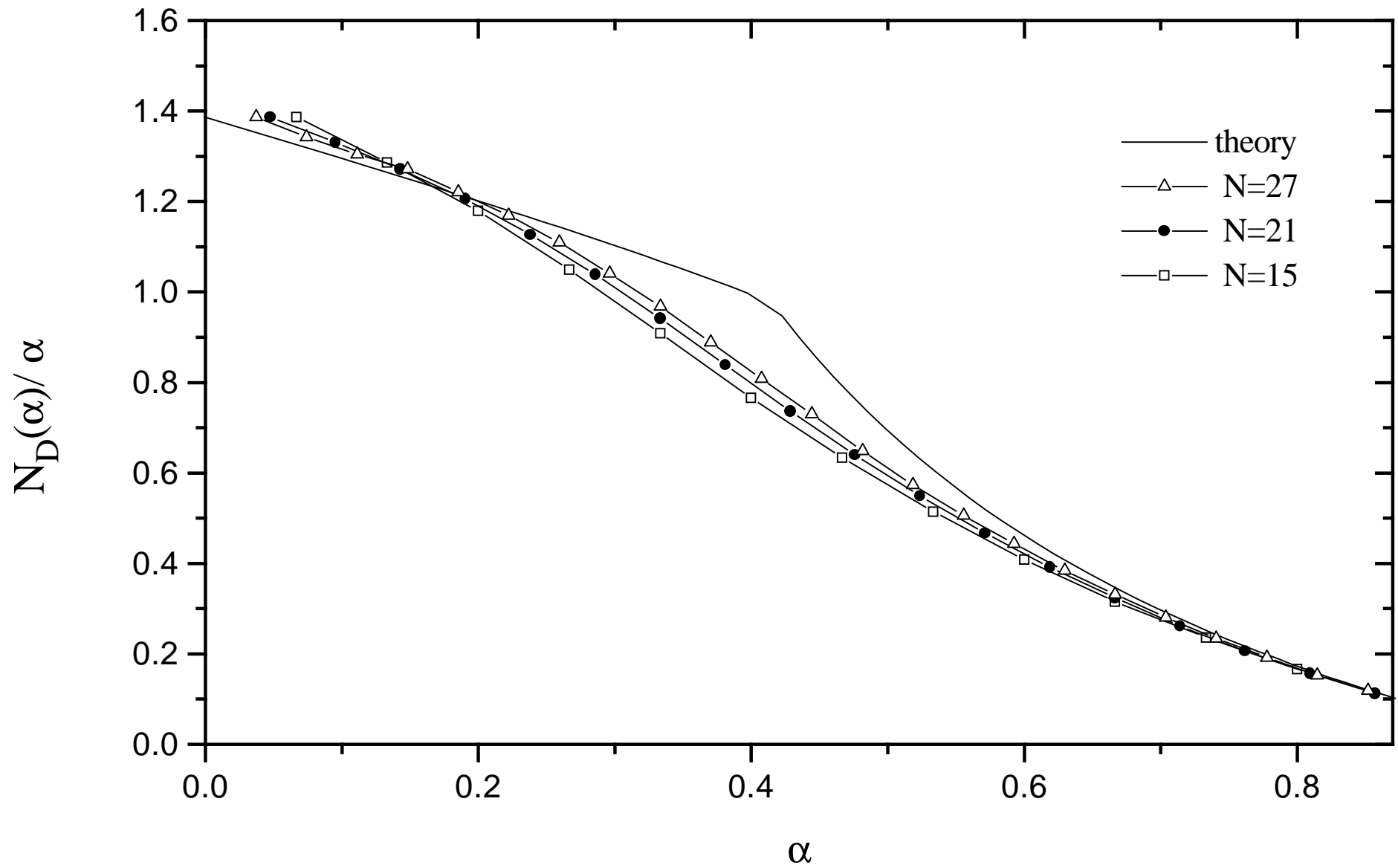
\begin{figure}
\caption{${\cal N}_D/\alpha$ versus $\alpha$ (theoretical continuous line and
numerical points).}
\label{fig8}
\end{figure}

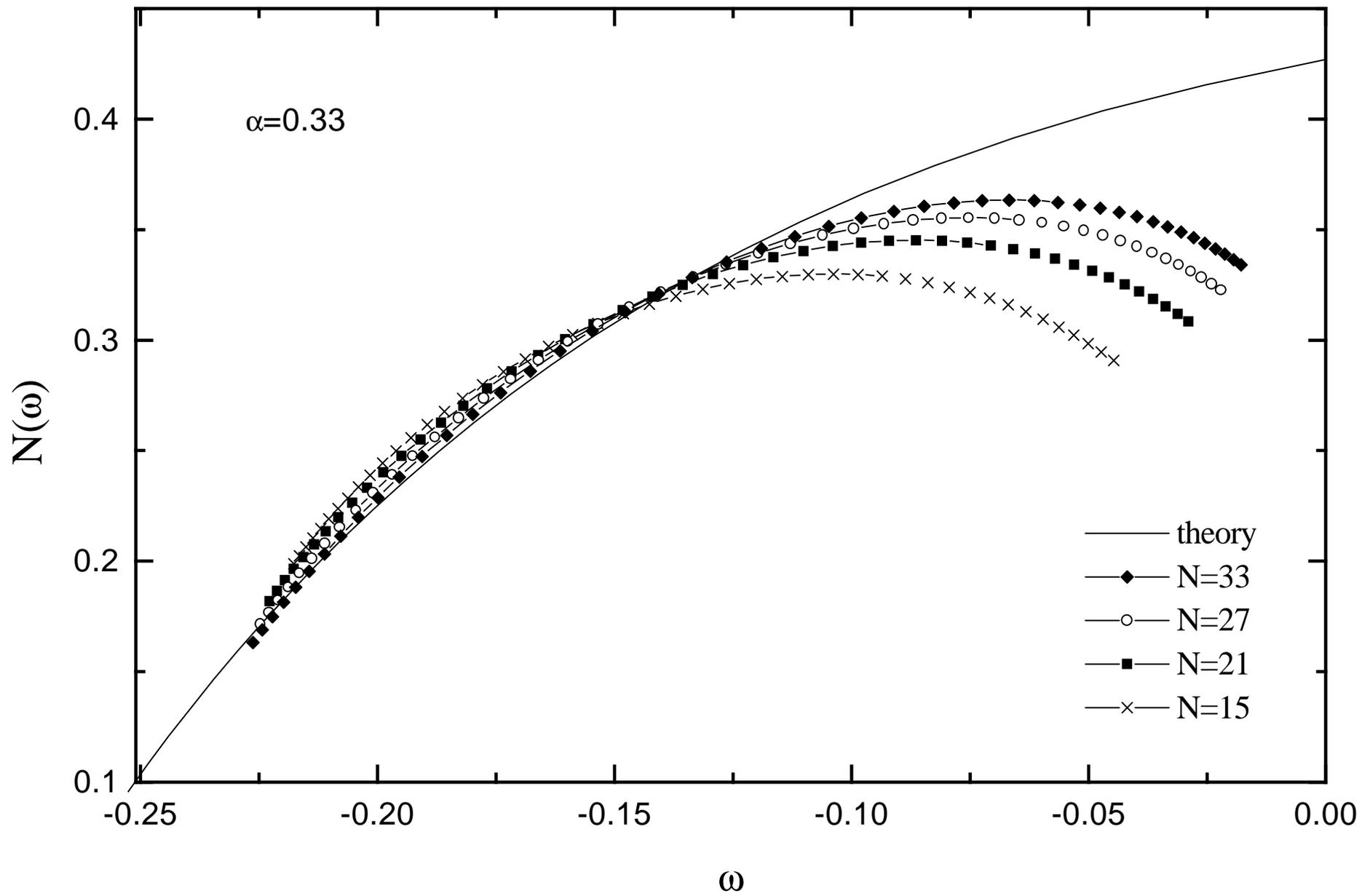
\begin{figure}
\caption{${\cal N}[w(r)]/\alpha $ versus $w(r)$ for fixed $\alpha=0.33$.}
\label{fig9}
\end{figure}

\end{document}